\documentclass[conference]{IEEEtran}
\IEEEoverridecommandlockouts
\usepackage{cite}
\usepackage{amsmath,amssymb,amsfonts}
\usepackage{algorithmic}
\usepackage{graphicx}
\usepackage{textcomp}
\usepackage{xcolor}
\usepackage{bm}
\def\BibTeX{{\rm B\kern-.05em{\sc i\kern-.025em b}\kern-.08em
    T\kern-.1667em\lower.7ex\hbox{E}\kern-.125emX}}
\begin{document}

\title{A Neural Network Detector for Spectrum Sensing under Uncertainties%\\
\thanks{This research was supported by the Data Science/Machine Learning Initiative of UC San Diego's ECE Dept.}
}

\author{\IEEEauthorblockN{Ziyu Ye}
\IEEEauthorblockA{\textit{Dept. ECE} \\
\textit{UC San Diego}\\
San Diego, United States \\
ziy076@kiwi-ml.ucsd.edu}
\and
\IEEEauthorblockN{Qihang Peng}
\IEEEauthorblockA{\textit{Sch. of Info. and Comm. Eng.} \\
\textit{UESTC}\\
Chengdu, China \\
anniepqh@uestc.edu.cn}
\and
\IEEEauthorblockN{Kelly Levick}
\IEEEauthorblockA{\textit{Dept. ECE} \\
\textit{UC San Diego}\\
San Diego, United States \\
klevick@ucsd.edu }
\and
\IEEEauthorblockN{Hui Rong}
\IEEEauthorblockA{\textit{Coll. of Info. and Comm. Eng.} \\
\textit{Harbin Engineering University}\\
Harbin, China \\
h1rong@eng.ucsd.edu}
\and
\IEEEauthorblockN{Andrew Gilman}
\IEEEauthorblockA{\textit{Sch. of Natural and Computational Sci.} \\
\textit{Massey University}\\
Auckland, New Zealand \\
A.Gilman@massey.ac.nz}
\and
\IEEEauthorblockN{Larry Milstein}
\IEEEauthorblockA{\textit{Dept. ECE} \\
\textit{UC San Diego}\\
San Diego, United States \\
milstein@ece.ucsd.edu}
\and
\IEEEauthorblockN{Pamela Cosman}
\IEEEauthorblockA{\textit{Dept. ECE} \\
\textit{UC San Diego}\\
San Diego, United States \\
pcosman@eng.ucsd.edu}
}

\maketitle

\begin{abstract}
Spectrum sensing is of critical importance in any cognitive radio system. When the primary user's signal has uncertain parameters, the likelihood ratio test, which is the theoretically optimal detector, generally has no closed-form expression. As a result, spectrum sensing under parameter uncertainty remains an open question, though many detectors exploiting specific features of a primary signal have been proposed and have achieved reasonably good performance. In this paper, a neural network is trained as a detector for modulated signals. The result shows by training on an appropriate dataset, the neural network gains robustness under uncertainties in system parameters including the carrier frequency offset, carrier phase offset, and symbol time offset. The result displays the neural network's potential in exploiting implicit and incomplete knowledge about the signal's structure.
\end{abstract}


\section{Introduction}
Cognitive radios (CRs) exploit under-utilized transmission opportunites in licensed communication systems. In the widely studied interweave CR paradigm \cite{AAli}, a secondary user (SU) can transmit in a frequency band only when no primary user (PU) is active in the band. To avoid collision with a PU, the SU senses the band periodically. In each period, the SU performs sensing in a small portion of the time (called a sensing interval) to detect the presence of any PU signal in the band. If no PU signal is detected, the SU transmits in the band until the next sensing interval comes; Otherwise, the SU stays silent. The detection accuracy in spectrum sensing is of critical importance for CR. According to detection theory, the optimal detector is the likelihood ratio test (LRT). In a CR system, however, the LRT is generally not applicable because the SU does not have perfect knowledge about the primary signal. With incomplete knowledge at the SU, the primary signal's parameters are uncertain. To perform LRT with uncertain parameters $\bm{\theta}$, one needs to average the conditional likelihood ratio over the probability density function (PDF) of $\bm{\theta}$. This in turn requires modeling the PDF of $\bm{\theta}$, and integrating the product of the PDF and the conditional likelihood ratio over $\bm{\theta}$. The former task is a high-dimensional estimation problem, which can be hard depending on the true underlying PDF of $\bm{\theta}$. The latter task is an integration which generally does not reduce to a closed form. One exception to this is the case in which the signal is narrow-band Gaussian noise. In this scenario, the likelihood ratio has a closed-form expression and is equivalent to the energy of the received waveform, for which the LRT reduces to energy detection. Since the narrow-band Gaussian can be seen as a distribution which provides no information about the signal's structure except its bandwidth, energy detection can be seen as the optimal detector for signals whose structure is completely unknown. In the more general case, the primary signal's structure is neither completely uncertain nor exactly known, and the LRT is not applicable because of the aforementioned reasons. This leaves how to utilize the incomplete knowledge about the primary signal an open question.       

Most conventional spectrum sensing algorithms can be seen as heuristic approaches of exploiting partial knowledge about the signal's structure. A typical strategy for doing this is feature matching: having prior knowledge about some feature(s) of the signal, the detector makes a decision based on whether the feature is present in the input. For example, \cite{PZhang} learns from data the leading eigenvector of the signal's time-domain covariance matrix as feature, and makes a decision by comparing it to the leading eigenvector of the sample covariance matrix computed from the input. The prior information required by a detector does not necessarily take explicit form. In \cite{YZeng}, the maximum-minimum-eigenvalue (MME) detection and the energy-with-minimum-eigenvalue (EME) detection make decisions based on the strength of the correlation observed in the input. Though no prior knowledge is required explicitly, MME and EME rely on the implicit assumption that the signal contains a relatively strong covariance structure. These conventional detection algorithms, though not theoretically optimal, are able to make use of incomplete knowledge about the primary signal and achieve better performance than energy detection in favorable circumstances. 

A neural network (NN) provides an alternative solution for the problem. A NN classifier is a numerical framework which can learn a discriminative model for a classification task through training on labeled data. In a binary classification task, if the cross-entropy loss is used, training encourages the NN to map the input to the log-posterior-probabilities of the two classes (ignoring an arbitrary constant term). Ideally, the difference between the NN's two outputs would converge to the log-likelihood ratio, which is the optimal test statistic for detection. As discussed earlier, the analytically derived likelihood ratio has no closed-form expression in the typical CR
environment. The NN can potentially solve this problem by learning a numerical approximation of the true underlying model. In addition, as a data-driven technique, the NN learns the model directly from data and thus avoids the parameter (distribution) estimation step required by the LRT.

The idea of applying a NN to spectrum sensing has been around for a few years. One popular application of the NN classifiers in spectrum sensing is spectrum prediction \cite{AEltholth,SBai,LYin,OWinston,RFleifel,AAgarwal,AShahid,MHuk,VTumuluru,NShamsi} (also called channel-occupancy prediction, or primary-user-activity prediction by some authors). In spectrum prediction, the NN is trained to predict the future occupancy state(s) from its history. The input feature is usually the occupancy states or the received power in the channel in a number of past sensing cycles. Another area where NNs found application in spectrum sensing is in cooperative sensing. In cooperative sensing, local decisions made by multiple SUs located at different geographic locations are combined (called decision fusion) to produce a decision of higher accuracy. In \cite{YChen}, a NN is used to facilitate the decision fusion by assessing the reliability (predicting the combining weight) of each local decision, based on the pattern formed by all local decisions. In these two applications, the NN is trained to exploit either temporal correlation or spatial correlation/diversity of the channel occupancy. The (temporally or spatially) local decisions, which serve as the input features, are usually produced by trivial detectors such as energy detection. While in these works the NN shows its effectiveness in exploiting ``high-level" information, the findings do not show whether a NN can be trained to exploit the internal structure of the signal.

In the literature, there are a handful of examples where a NN is trained to exploit the primary signal's internal structure. In \cite{YTang}, a NN takes the energy and three cyclostationary features of the received waveform as input and detects the presence of an AM signal in AWGN. In \cite{JPopoola}, the author performs spectrum sensing via modulation classification. A NN takes 8 statistical features as input, and outputs a length-13 vector corresponding to 12 possible modulation schemes and the idle-state. The decision is made according to the largest entry in the vector. Though not in the scope of spectrum sensing, in \cite{AFehske}, a NN is trained to recognize cyclostationary features for the purpose of modulation classification. In \cite{NFarsad}, a sliding-bidirectional-recurrent-neural-network (SBRNN) is trained as an intended receiver. This paper also proposes using NN as a solution to the parameter uncertainty problem in communication. In most of these works, the NN is used as one step in detection. Usually a preprocessing step is used to produce some analytically designed features, and the NN maps the feature space to a final test statistic. While analytically designed feature-extracting preprocesses may help reduce the input dimensionality of the NN, they could cause information loss as well. We hypothesize that, given abundant labeled data, an NN trained as an end-to-end detector can approach the theoretically optimal detector, which gives it an advantage over any NN detector based on engineered features.

In this work, a NN is trained as a detector which takes raw samples from the baseband waveform as input and produces a test statistic to decide whether the primary signal is present or absent in the waveform. To demonstrate the NN's potential in detecting signals with parameter uncertainty, we consider an unknown offset in the carrier frequency of the primary signal. The goal is to train the NN so that it is robust under this uncertainty. Random carrier phase and symbol timing of the primary signal are also considered. In Section II, we describe the system model and introduce three scenarios covering different possibilities in inter-carrier-interference (ICI). In Section III, experimental results are presented and discussed.  

\section{System Model}
Consider an opportunistic spectrum access scenario where an SU performs narrow-band detection to utilize a certain subcarrier in a multi-tone primary communication system. The subcarriers of the primary system are separated by guardbands, and each carries a modulated signal with parameters listed in Table \ref{tab1}. Only the downlink transmission is considered, and all subcarriers are assumed to have the same signal power when active. The SU listens to the channel for a sensing interval and makes a decision on whether the PU is active in the subcarrier of interest. The received waveform is first passed through a bandpass filter (BPF) to eliminate noise beyond the signal's bandwidth, then shifted to baseband, sampled, and passed to the decision-making algorithm. Although the primary system described here has multiple subcarriers, the focus of this paper is on using the received waveform within a single subcarrier's bandwidth to make a sensing decision.

\begin{table}[htbp]
\caption{System Parameters}
\begin{center}
\begin{tabular}{|c|l|}
\hline
\multicolumn{2}{|c|}{\textbf{Parameters of the Primary Signal}} \\
\hline
\textit{Carrier phase} & Random, uniformly distributed in range $[0,2\pi)$\\
\hline
\textit{Symbol Time} & Random, uniformly distributed in range $[0,T_{sym})$, \\ 
\textit{Offset}& where $1/T_{sym}$ is the symbol rate\\
\hline
\textit{Pulse Shape} & Root-raised-cosine (RRC) pulse \\
& with roll-off factor 0.35\\
\hline
\textit{Modulation} & QPSK\\
\hline
\textit{SNR} \textit{(Post-LPF)} & 0 dB when no center frequency offset exists \\
\hline
\multicolumn{2}{|c|}{\textbf{Sensing Parameters}} \\
\hline
\textit{Sensing Duration} & $11T_{sym}$ \\
\hline
\textit{Sample Rate} & $10/T_{sym}$ \\
\hline
\end{tabular}
\label{tab1}
\end{center}
\end{table}

In narrow-band detection, ideally, the BPF's pass-band should match the primary signal's band exactly, so that the SNR is maximized if the signal is present. The match, however, can be inexact due to either the SU's inaccurate knowledge about the primary system or Doppler shift caused by motion of the user. In this work, we consider an offset between the primary signal's carrier frequency and the center frequency of the BPF used by the SU. Our main goal is to train the NN as a decision-making algorithm resilient to the frequency offset. The NN's robustness under random carrier phase and symbol timing is also considered.

The effect of the frequency offset depends on the guard band. When the frequency offset is smaller than the guard band's bandwidth, the frequency offset causes only pulse-shape distortion. The pulse shape is distorted because a part of the pulse in the frequency domain is cut off by the BPF, which also causes the loss of a part of the signal power. With frequency offset, carrier is not entirely removed when the signal is brought to baseband, which also contributes to pulse-shape distortion. If the frequency offset exceeds the guard bandwidth, ICI occurs in addition to the aforementioned effects. In this case, the received waveform contains up to two trains of distorted root-raised-cosine (RRC) pulses, which have independent carrier phases and symbol time offsets, and carry independent source data. In this work, instead of assuming a specific bandwidth for the guard band, two extreme scenarios are studied: one has no ICI despite the frequency offset, and the other has no guard band so any non-zero frequency offset causes ICI. These two scenarios can be seen as representing the two cases in which the frequency offset falls below or beyond the guard band's bandwidth, and the results obtained can be generalized to any scenario with a realistic guard band.  

To simulate the scenario where ICI may occur, it is required to specify the correlation between the presence of the signal and that of the ICI. The correlation between the primary activities on two adjacent subcarriers depends on both the channel allocation and the type of PU activity. For example, if the two subcarriers are allocated to two different PUs, their activities are likely independent; if two carriers are allocated to the same PU, their activities can be correlated. We again consider two extreme scenarios. In one scenario, the signal and ICI are either both present or both absent. In the other, the presences of the signal and the ICI are independent. In summary, three scenarios are studied in this work:
\begin{itemize}
\item Scenario A : No ICI
\item Scenario B : Signal and ICI occur together
\item Scenario C : Signal and ICI occur independently
\end{itemize}

\section{Experiments and Discussion}
\subsection{Experimental Setup}
A fully-connected feed-forward NN with two hidden layers (400 nodes each, ReLU activation functions) is used as a binary classifier for PU detection. The input consists of the samples collected during one sensing interval, with real and imaginary parts of each sample treated as separate real features. Training was performed using cross-entropy loss, Adam optimizer with initial learning rate of 5E-04, and batch size of 1E+03. Early termination at lowest validation loss was used in all training runs. Training (8E+6 examples), validation (1E+5 examples) and test (1E+6 examples) sets consisted of software-simulated data generated by GNU Radio. 

\begin{table}[htbp]
\caption{Hyperparameters of the NN and its training}
\begin{center}
\begin{tabular}{|c|p{2in}|}
\hline
\multicolumn{2}{|c|}{\textbf{NN Specification}} \\
\hline
\textit{Input Size} & 222 (twice the number of samples in the sensing interval)\\
\hline
\textit{Output Size} & 2 (correponding to the two classes)\\
\hline
\textit{Hidden Layers} & 2 fully-connected layers, each of size 400\\
\hline
\textit{Nonlinearity} & ReLU\\
\hline
\multicolumn{2}{|c|}{\textbf{Training Settings}} \\
\hline
\textit{Loss Function} & Cross Entropy Loss\\
\hline
\textit{Optimizer} & Adam\\
\hline
\textit{Learning Rate} & 5E-04\\
\hline
\textit{Batch Size} & 1E+03 examples\\
\hline
\textit{Epochs} & Not fixed, early termination was used.\\
\hline
\textit{Dateset Size} & 8E+06 examples (training), 1E+05 examples (validation), 1E+06 examples (test)\\
\hline
\end{tabular}
\label{tab2}
\end{center}
\end{table}

Because an SU typically performs noncoherent and asynchronous detection, the received signal is assumed to contain a random carrier phase offset $\theta \sim \bm{U}(0,2\pi)$ and a random time offset $\tau \sim \bm{U}(0,T_{sym})$, where $\bm{U}(a, b)$ denotes a uniform distribution over the interval $(a,b)$ (see Table \ref{tab1} for all system parameters). A NN  trained on a dataset generated with random $\theta$ and $\tau$, is expected to be insensitive to changes in $\theta$ and $\tau$. Our first experiment verifies this by comparing such a robust NN with a baseline NN, trained on data with no carrier phase offset ($\theta=0$) and no time offset ($\tau=0$).

In the second set of experiments, we investigate robustness against carrier frequency offset. The NN is trained on mixed datasets, which consist of equal numbers of examples generated for the three scenarios. This is because the NN detector needs to handle uncertain scenarios in practice: a typical SU has no prior knowledge on whether the ICI exists, or on the correlation between the signal and the ICI. To gain resilience to the unknown frequency offset, each example in a training set has a random frequency offset (as well as a random phase offset and timing offset as in experiment 1). We investigate two different distributions for the frequency offset in training: $\bm{U}(-W/2, W/2)$, $\bm{U}(-W/4, W/4)$, where $W$ denotes the signal's bandwidth. These robust NNs are compared to a baseline, which has been trained on data with no frequency offset (only random phase and timing offsets). The robust and baseline NNs are tested in each of the three scenarios separately. In each scenario, the NNs are tested on frequency-offset-specific test sets, covering frequency offsets ranging from $-W/2$ to $W/2$, with step size $W/20$. The performance metric is the detection probability with the false alarm probability fixed at 1\%. We observed all ROC curves and detection probability at other false alarm rates have similar relative results and are omitted for brevity. For comparison, energy detection is also tested. 

In summary, 4 experiments are conducted to investigate the NN's robustness under uncertainties in the primary signal's carrier phase,  symbol timing and carrier frequency. Goals of the experiments and the NNs compared in the experiments are listed in Table \ref{tab_exp}.
\begin{table}[htbp]
\caption{Experiments and NNs}
\begin{center}
\begin{tabular}{|c|p{2.5in}|}
\hline
\multicolumn{2}{|c|}{\textbf{Experiments}} \\
\hline
\it{Name} & \multicolumn{1}{|c|}{\it{Goal of the Experiment}}\\
\hline
I & Compare the sensitivities of NN1 and NN2 to the carrier phase offset $\theta$ and the time offset $\tau$. \\
\hline
II.A & Compare the resiliences of NN2, NN3, NN4 to frequency offsets. Scenario A is considered.\\
\hline
II.B & Compare the resiliences of NN2, NN3, NN4 to frequency offsets. Scenario B is considered.\\
\hline
II.C & Compare the resiliences of NN2, NN3, NN4 to frequency offsets. Scenario C is considered.\\
\hline
\multicolumn{2}{|c|}{\textbf{Neural Networks}} \\
\hline
\it{Name} & \multicolumn{1}{|c|}{\it{Training Condition}}\\
\hline
NN1 & No frequency offset, \\
& $\theta = 0$, $\tau=0$\\
\hline
NN2 & No frequency offset, \\
& $\theta \sim \bm{U}(0,2\pi)$, $\tau \sim \bm{U}(0, T_{sym})$\\
\hline
NN3 & Random frequency offsets following $\bm{U}(-W/4,W/4)$,\\
& $\theta \sim \bm{U}(0,2\pi)$, $\tau \sim \bm{U}(0, T_{sym})$\\
\hline
NN4 & Random frequency offsets following $\bm{U}(-W/2,W/2)$, \\
& $\theta \sim \bm{U}(0,2\pi)$,  $\tau \sim \bm{U}(0, T_{sym})$\\
\hline
\end{tabular}
\label{tab_exp}
\end{center}
\end{table}

%\begin{table}[htbp]
%\caption{Hyperparameters of the NN and its training}
%\begin{center}
%\begin{tabular}{|c|p{2in}|}
%\hline
%\multicolumn{2}{|c|}{\textbf{NN Specification}} \\
%\hline
%\textit{Input Size} & 222 (twice the number of samples in the sensing interval)\\
%\hline
%\textit{Output Size} & 2 (correponding to the two classes)\\
%\hline
%\textit{Hidden Layers} & 2 fully-connected layers, each of size 400\\
%\hline
%\textit{Nonlinearity} & ReLU\\
%\hline
%\multicolumn{2}{|c|}{\textbf{Training Settings}} \\
%\hline
%\textit{Loss Function} & Cross Entropy Loss\\
%\hline
%\textit{Optimizer} & Adam\\
%\hline
%\textit{Learning Rate} & 5E-04\\
%\hline
%\textit{Batch Size} & 1E+03 examples\\
%\hline
%\textit{Epochs} & Not fixed, early termination was used.\\
%\hline
%\textit{Dateset Size} & 8E+06 examples (training), 1E+05 examples (validation), 1E+06 examples (test)\\
%\hline
%\end{tabular}
%\label{tab2}
%\end{center}
%\end{table}

\subsection{Results and Discussion}
\subsubsection{Experiment I}
\begin{figure}[htbp]
\centerline{\includegraphics[trim=0 0 0 20,clip,width=3.7in]{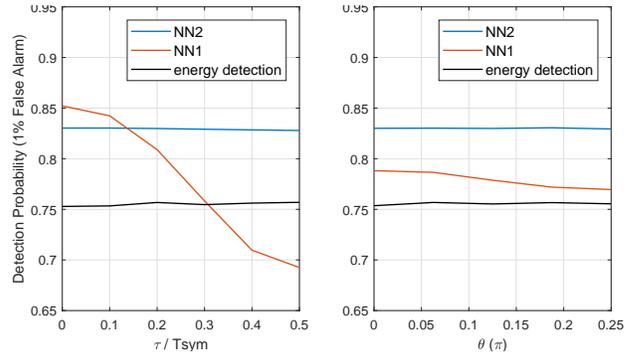}}
\caption{NNs' sensitivity to carrier phase offset and time offset} 
\label{fig_phase_timing_2}
\end{figure}

NN1, NN2, and an energy detector are tested across time offset $\tau$ in the range $(-T_{sym}/2, T_{sym}/2)$. The testsets contain random carrier phases ($\theta\sim \bm{U}(0,2\pi)$). The results are plotted in Fig. \ref{fig_phase_timing_2} (left). Only the positive range of $\tau$ is shown as the curves are roughly symmetrical about $\tau=0$. The performance of NN1 can be observed to drop quickly with an increase in $\tau$. On the contrary, NN2 produces roughly the same performance across all $\tau$ values tested. The three detectors are also tested across carrier phase offset $\theta$ in the range $(-\pi/4, \pi/4)$. The results are shown in Fig. \ref{fig_phase_timing_2} (right). The testsets contain random time offsets ($\tau\sim\bm{U}(0, T_{sym})$). Similarly, NN2 appears more robust than NN1. It can also be observed that NN1 is more sensitive to changes in $\tau$ than changes in $\theta$. These results show that a NN can acquire robustness under carrier phase offset and symbol timing offset through training on random $\theta$ and $\tau$. The performance of the energy detector is insensitive to changes in $\tau$ and $\theta$; however, it is not able to achieve the same performance as NN2.

\subsubsection{Experiment II.A}
\begin{figure}[htbp]
\centerline{\includegraphics[trim=0 0 0 20,clip,width=3.7in]{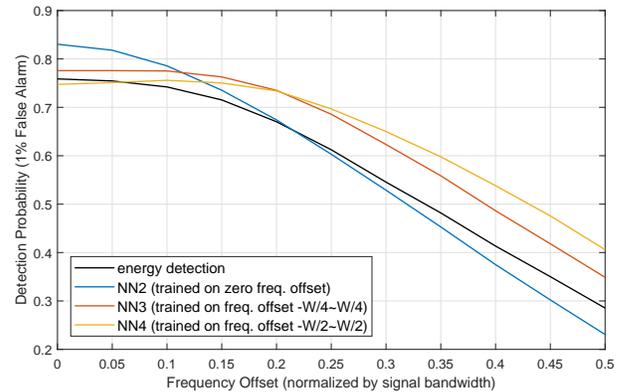}}
\caption{Detection probability (with 1 percent false alarm) versus carrier frequency offset, no ICI.}
\label{fig_no_ici}
\end{figure}

Performances of the detectors in Scenario A are shown in Fig.~\ref{fig_no_ici}. With ICI absent, a frequency offset leads to distortion of the pulses in the primary signal, and consequentially loss of SNR. The SNR loss is reflected by the performance of the energy detection, which drops when the frequency offset increases. The SNR loss is accompanied by a loss of information in the received waveform, to which no detector can be resilient. It is possible, however, to have a detector relatively insensitive to pulse-shape distortion. In Fig.~\ref{fig_no_ici}, a detector's resilience to pulse-shape distortion is reflected by its performance gain over energy detection at higher frequency offsets. NN2 performs better than other detectors in its training condition. However, it is not resilient to frequency offsets: its performance gain over energy detection is quickly lost when the frequency offset increases. NN3 trained on $\bm{U}(-W/4, W/4)$ exhibits more robustness. It performs uniformly better than the energy detector in the range of frequency offsets plotted. NN4 trained on $\bm{U}(-W/2, W/2)$ performs even better at large frequency offsets, but its performance at no frequency offset is worse than the energy detection. Clearly, no single detector is uniformly the best across all frequency offsets. Training on a wider range of frequency offsets makes the detector more resilient to large frequency offsets, at the cost of a performance loss at low frequency offsets.        

\subsubsection{Experiment II.B}
\begin{figure}[htbp]
\centerline{\includegraphics[trim=0 0 0 20,clip,width=3.7in]{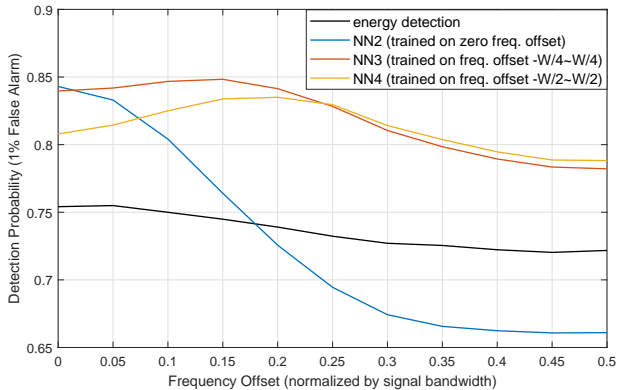}}
\caption{Detection probability (with 1 \% false alarm) versus carrier frequency offset, ICI occurs together with signal.}
\label{fig_ici}
\end{figure}

The effect of ICI whose presence is strongly correlated with that of the signal is examined in Scenario B. The detectors' performances in Scenario B are shown in Fig.~\ref{fig_ici}. A similar trend exists that training on a wider range of frequency offsets leads to more robustness but somewhat worse performance at low frequency offsets. Despite the similarity, a few differences from Scenario A can be observed in  Fig.~\ref{fig_ici}. First, the energy detector's performance drops more slowly with frequency offset increase in Fig.~\ref{fig_ici} than in Fig.~\ref{fig_no_ici}. This is reasonable because the SNR loss is less severe: In Scenario B, the target of detection is the sum of the signal and the ICI. When frequency offset increases, the received signal power decreases, but ICI power increases. Another difference from Scenario A is that, at large frequency offsets, the performance of NN2 is notably worse than that of energy detection. This indicates that NN2 is inefficient in making constructive use of the ICI, probably because of the absence of ICI in its training. The other two NNs, trained on datasets that contain ICI, are able to maintain their performance gains over the energy detector.

\subsubsection{Experiment II.C}
\begin{figure}[htbp]
\centerline{\includegraphics[trim=0 0 0 20 ,clip,width=3.7in]{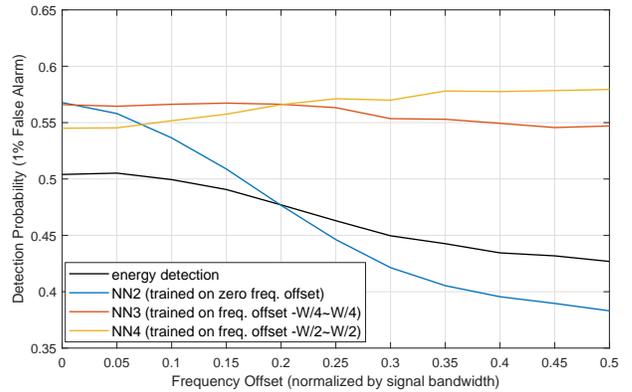}}
\caption{Detection probability (with 1 \% false alarm) versus carrier frequency offset, signal and ICI occur independently.}
\label{fig_ici_indept}
\end{figure}

Scenario C is the same as Scenario B, except that the presence of ICI is independent of the presence of signal. In Scenario C, the labelling of training and test data deserves special note. Because the purpose of spectrum sensing is to avoid collision between the SU and PU(s), when the SU's sensing/transmission band overlaps with two adjacent bands in the primary communication system, the SU should avoid collision with the primary signal in either of the two bands. Based on this consideration, an example is labeled as idle only if both signal and ICI are absent. Fig.~\ref{fig_ici_indept} shows the performance of the detectors in Scenario C.  Similar to the observation made in the other two scenarios, the NNs trained on a wider range of frequency offsets are more resilient to large frequency offsets, but perform worse at zero frequency offset. One may note that because of the labelling rule, the energy detection's performance at zero frequency offset in Fig.~\ref{fig_ici_indept} is different from that in Fig.~\ref{fig_no_ici} and Fig.~\ref{fig_ici}.

\section{Conclusion}
We trained a NN as a detector resilient to offsets in the primary signal's carrier phase, symbol timing, and carrier frequency. Experimental results show that robustness under uncertain carrier phase and symbol timing can be obtained by training the NN on data containing random carrier phases and time offsets. By training on a dataset containing mixed frequency offsets, the NN can obtain some resilience to the pulse-shape distortion caused by the frequency offset, at the cost of a slight decrease in performance at zero frequency offset. The trade off between the resilience to frequency offset and the performance at no frequency offset can be adjusted by changing the distribution of the frequency offset in the training set.      

%\section*{Acknowledgment}

\end{document}